\documentclass[aps,twocolumn,amsfonts,showpacs,superscriptaddress]{revtex4-1}
\usepackage{amssymb,amsmath,amsthm,booktabs,mathtools}
\usepackage{bm}
\usepackage[dvipsnames]{xcolor}
\usepackage{tikz}
\usepackage{enumitem}
\usepackage{braket}
\usepackage{gensymb}
\usepackage{amsmath}
\usepackage{cancel}
\usepackage{soul} 

\usepackage[separate-uncertainty = true,multi-part-units=single]{siunitx} 

\usepackage[version=4]{mhchem}

\usepackage{braket}

\begin{document}

\title{$\Lambda$-enhanced gray-molasses loading and EIT cooling of neutral atoms in nanophotonic traps}

\author{Lucas Pache}
\email{pachluca@hu-berlin.de}
\author{Antoine Glicenstein}
\altaffiliation{Present address: LP2N, Laboratoire Photonique, Numérique et Nanosciences, Université de Bordeaux-IOGS-CNRS: UMR 5298, F-33400 Talence, France}
\author{Philipp Schneeweiss}
\author{Jürgen Volz}
\author{Arno Rauschenbeutel}
\email{arno.rauschenbeutel@hu-berlin.de}
\author{Riccardo Pennetta}
\affiliation{Department of Physics, Humboldt-Universität zu of Berlin, 10099 Berlin, Germany}

\begin{abstract}
Nanophotonic traps for cold atoms typically have trap volumes that are orders of magnitude smaller than, e.g., free-space optical tweezers. This makes efficient loading of these traps challenging, thereby limiting the total number of atoms coupled to the nanophotonic waveguide. Here, we demonstrate that $\Lambda$-enhanced gray-molasses ($\Lambda$GM) can substantially increase the number of trapped atoms in a nanofiber-based cold-atom setup. Specifically, we observe a six-fold increase in the number of loaded atoms compared to conventional red-detuned polarization gradient cooling. Despite the unusually small depth of our optical trap of only 24 $\mu$K, we load about 4000 individual Cesium atoms, achieving optical depths exceeding 140 and reaching the collisional blockade regime over a length of approximately 1 mm. After loading, we perform efficient EIT-assisted cooling that is found to increase the trap storage time to 400(9) ms. This is a 5-fold improvement over the passive storage time. Remarkably, EIT-cooling also works with two co-propagating nanofiber-guided light fields and requiries only about a few hundred picowatt of optical power. Our results provide an efficient method to boost both the number of loaded atoms and the storage time of nanophotonic atom traps.
\end{abstract}

\maketitle

\section{Introduction}

Interfacing trapped neutral atoms with nanophotonic waveguides enables the experimental observation of stable and efficient interactions between guided photons and ensembles of up to thousands of atoms~\cite{Chang2018, Sheremet2023, GonzalezTudela2024, Li_2024}. In the last decade, this capability has enabled significant advances in, e.g.,  quantum optics~\cite{Sorensen2016, Corzo2016, Lodahl2017, Nayak2019, Prasad2020, Pucher2022, Vylegzhanin2023}, quantum information processing~\cite{Sayrin2015, Gouraud2015, Corzo2019, Scheucher2016, Tiecke2014, Dordevic2021, Shomroni2014,  Rosenblum2015a}, and studies of collective radiative phenomena~\cite{Goban2015, corzo_large_2016, Ruddell_2017, Kato_2019, liedl_observation_2023}. These achievements have been made possible by the development a wide range of nanophotonic platforms for cold atom experiments, including optical nanofibers~\cite{Vetsch2010, Goban2012}, slow-light waveguides~\cite{Goban2014}, whispering-gallery-mode resonators~\cite{Scheucher2016, Alton2010, Zhou2024a, Shomroni2014}, and photonic crystal cavities~\cite{Thompson2013, Menon2024}.

In all these different platforms, atoms are typically trapped either by optical potentials formed by guided modes~\cite{Vetsch2010} or by external optical tweezers, which are partially reflected by the waveguide \cite{Thompson2013}. Both approaches share a defining feature: the trapping potentials occupy much smaller volumes compared to conventional atom traps in free space. Specifically, typical volumes for nanophotonic traps are a few orders of magnitude smaller than $\lambda^{3}$~\cite{Chang2018}. This severely renders the loading from a thermal cloud of laser-cooled atoms challenging. As a consequence, efficient and reliable controlled atom loading remains a central challenge in this field. 

Recently, $\Lambda$-enhanced gray molasses ($\Lambda$GM) cooling~\cite{boiron_three-dimensional_1995, hsiao__2018, grier_ensuremathlambda-enhanced_2013, Gruenzweig2010, Wang2022} has emerged as a highly effective technique for achieving high loading efficiencies in optical dipole traps for neutral atoms~\cite{lester_rapid_2015, brown_gray-molasses_2019, Glicenstein2021}. By combining velocity-selective coherent population trapping with polarization-gradient cooling in a blue-detuned configuration, $\Lambda$GM realizes sub-Doppler cooling while strongly suppressing photon scattering, yielding temperatures on the order of $1~\mu$K for Cesium atoms~\cite{hsiao__2018}. \\ \indent
In this work, we experimentally demonstrate that $\Lambda$GM and the closely related technique of electromagnetically induced transparency (EIT) cooling~\cite{Morigi2000, Roos2000, Kampschulte2014, Haller2015, Chow2024} can address several of the challenges posed by nanophotonic interfaces for laser-cooled atoms.
Specifically, we show that $\Lambda$GM can substantially enhances the loading of nanofiber-based traps for cold atoms. In our experiment, we observe a six-fold increase in the number of trapped atoms compared to red-detuned polarization-gradient cooling (RPGC). This enables the trapping of about 4000 individual atoms and reaching maximum ocupation in the the collisional blockade regime~\cite{schlosser_collisional_2002-1}. 
Using the same optical beams employed for $\Lambda$GM, we implement EIT–assisted cooling of the atom trapped near the nanofiber. This is particularly noteworthy because our experiment uses very shallow traps with a trap depth of only 24~$\mu$K, for which standard RPGC was found to be ineffective. This approach not only yields a 5-fold increase in the atom storage time, but also allows us to preserve the comparatively high atom–photon interaction strength observed immediately after loading over very long storage times. Finally, we demonstrate that EIT-cooling can also be implemented by exclusively using guided beams with a total power of just a few hundred pW. \\ \indent
\section{Experimental setup}
\begin{figure}
	\centering
	\includegraphics[width=0.5\textwidth]{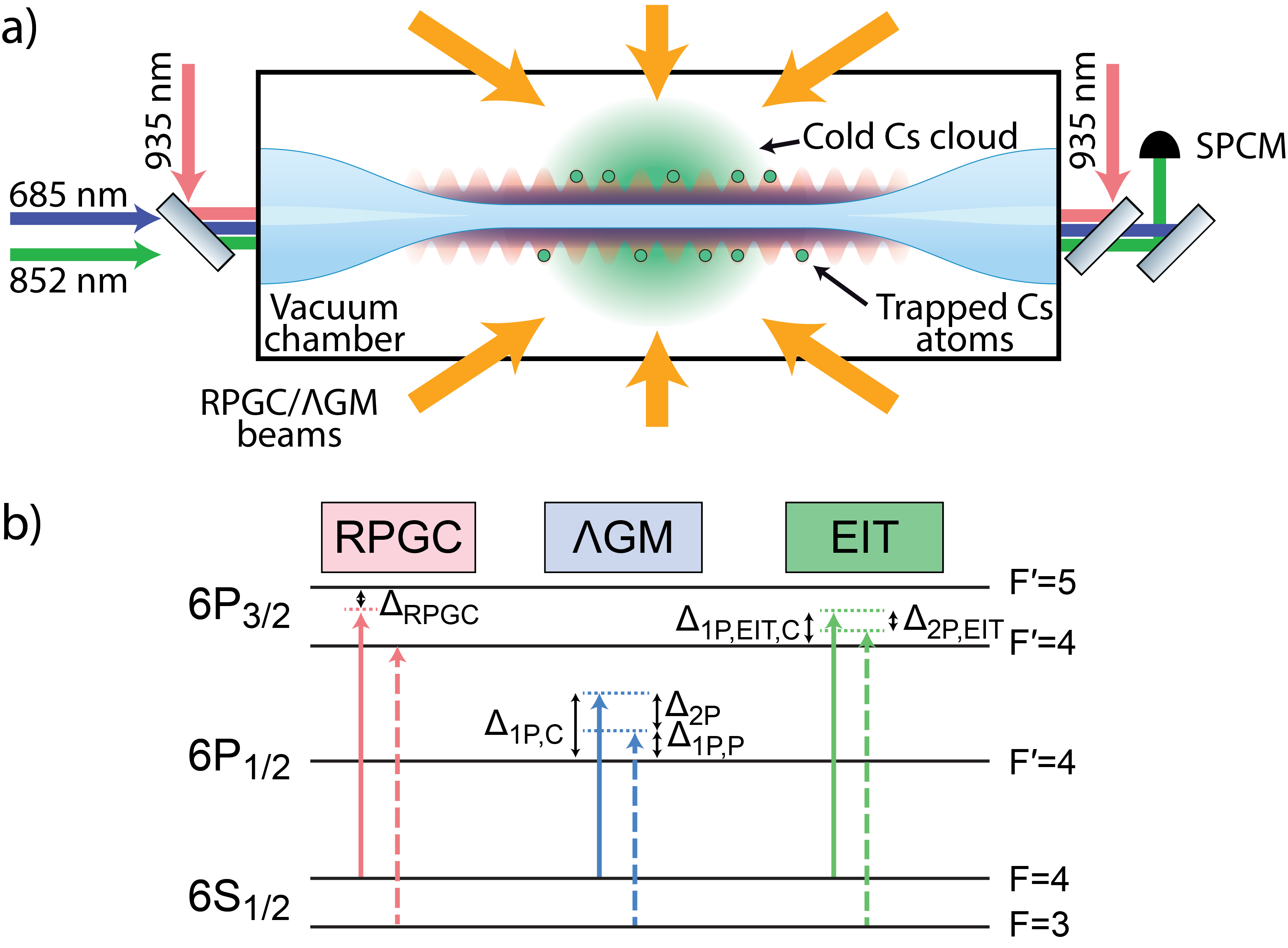}
	\caption{ a) Experimental setup for our nanofiber-based cold atom interface. We prepare a cloud of cold cesium atoms close to a $\SI{225}{\nm}$-waist-radius nanofiber by using a 3D-MOT and molasses configuration with $\sigma^{+} $and $\sigma^{-}$ polarization of the counter-propagating light fields. We trap the atoms close to the surface of the nanofiber using a fiber-guided red-detuned light field $(\lambda_\mathrm{red} = \SI{935}{\nm}$) in a standing wave configuration and a fiber-guided blue-detuned light field ($\lambda_\mathrm{blue} = \SI{685}{\nm}$) in a running wave configuration. We use fiber-guided light fields on the D2 transition ($\lambda_\mathrm{green} = \SI{852}{\nm}$) to probe the atoms and to perform EIT-cooling. b) Energy level scheme of the transitions used for molasses cooling: The RPGC-molasses is performed on the D2 cycling transition. The $\Lambda$GM-cooling is performed on the D1-transition such that the main cooler light field and the phase-coherent second light field couple the two hyperfine ground states to form a $\Lambda$-system via the $F' = 4$ hyperfine exited state. The $\Lambda$-system to conduct EIT-cooling is realized on the D2 line via the $F' = 4$ hyperfine exited state.
    }
    \label{fig:Setup}
\end{figure}
The experimental setup is shown schematically in Fig.~\ref{fig:Setup}a). We interface an ensemble of laser-cooled Cesium atoms with the evanescent field of light propagating in an optical nanofiber with a nominal radius of 225 nm. The latter is realized as the waist of a tapered optical fiber. The atoms are trapped in the evanescent field of the nanofiber using a two-color dipole trap~\cite{Vetsch2010}. In our experiment, the red- and blue-detuned trapping light fields operate at wavelengths of $\lambda_{red}$~=~935 nm and $\lambda_{blue}$~=~685 nm, which is close to magic wavelengths of the D2-transition~\cite{Goban2012, le_kien_dynamical_2013}. For all results in this paper, the red-detuned trapping field is set to a power of $ P_\mathrm{red} = \SI{55}{\uW}$ and is launched from both ends of the nanofiber to create a standing-wave dipole trap. The blue-detuned beam has a power of $P_\mathrm{blue}=\SI{7.75}{\mW}$. This configuration results in traps with a calculated depth of 24 $\mu$K with a minimum located about 270 nm from the nanofiber surface~\cite{pennetta2025hybridtrappingcoldatoms}. Due to their small trapping volumes of about \SI{1e-16}{\centi\meter\cubed}, the nanofiber traps have an occupancy of at most one atom per trapping site in the collisional blockade regime~\cite{schlosser_collisional_2002-1, vetsch_optical_2010}. The experiment begins by preparing a cloud of laser-cooled atoms in a magneto-optical trap. We then proceed with an initial stage of RPGC on the D2 transition (see Fig.~\ref{fig:Setup}b), yielding an atomic temperature of \SI{5.5(2)}{\micro\K}, that was measured via time-of-flight imaging. Subsequently, we implement $\Lambda$GM cooling~\cite{grier_ensuremathlambda-enhanced_2013, hsiao__2018}. In this stage, two phase-coherent lasers in a $\Lambda$ configuration are applied on the D1 transition, coupling the two hyperfine ground states to the $6\mathrm{P}_{1/2},\,F' = 4$ excited state. The stronger of the two fields is blue-detuned by $\Delta_{\mathrm{1P}} = 4.5\,\Gamma_{\mathrm{D1}}$ from the $F = 4 \rightarrow F' = 4$ transition. The two-photon detuning is defined as $\Delta_{\mathrm{2P}} = \Delta_{\mathrm{1P,P}} - \Delta_{\mathrm{1P,C}}$ (see Fig.~\ref{fig:Setup} b)). The frequency of the phase coherent second field is used to control the two-photon detuning, $\Delta_{\mathrm{2P}}$, of the $\Lambda$ scheme (see Fig.~\ref{fig:Setup}b). After \SI{150}{\ms} of $\Lambda$GM cooling, the atomic temperature was found to be \SI{1.9(1)}{\micro\K} (see the Supplementary Material for more details of experimental sequence and the $\Lambda$GM cooling scheme). We estimate the optical depth (OD) of the nanofiber-trapped ensemble by performing transmission spectroscopy through the nanofiber on the $F = 4 \rightarrow F' = 5$, D2 transition. In addition, we  estimate the total number of trapped atoms, $N$, using the technique described in Ref.~\cite{Beguin2014}. This method consists of measuring the optical energy required to pump an ensemble initially prepared in the $F = 4$ ground state into the $F = 3$ ground state, using a laser resonant with the $F = 4 \rightarrow F' = 4$, D2 transition. With these two independent measurements, we can estimate the atom-photon coupling strength $\beta$, defined as the probability for an excited atom to decay by emitting a photon into the waveguide, as $\beta = \mathrm{OD}/(4N)$. Immediately after loading, we measure $\beta =$ \SI{1.08(0.03)}~\%. \\

\section{Loading of nanofiber-based trap}
\begin{figure}
	\centering
	\includegraphics[width=0.44\textwidth]{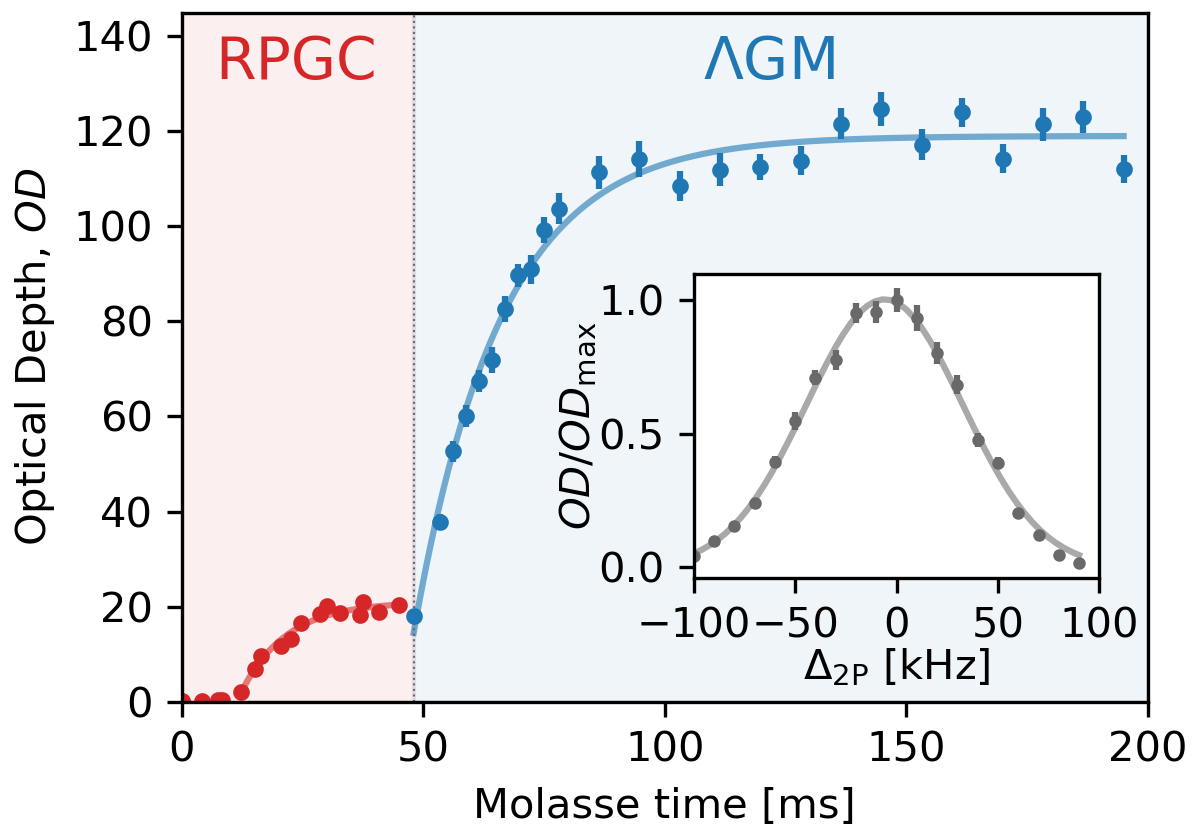}
	\caption{Measurement of the OD of the trapped atomic ensemble for different loading times. During the red-shaded time interval, the atoms are loaded using the RPGC-molasses for up to \SI{50.0}{ms} when they reach a steady state OD of \SI{20.4(1.2)}~. During the blue-shaded time interval, the atoms are loaded further using the $\Lambda$GM-molasse. During this second stage, we measure an increase in the OD by a factor of six with a maximum of \SI{125(4)}~. Lines: Exponential fits to the data. Inset: Measurement of the final OD in relation to the maximally measured OD for different two-photon detuning $\Delta_\mathrm{2P}$. The dependence is well described by a Gaussian distribution with a maximum close to the two-photon resonance (grey line).}
    \label{fig:Loading}
\end{figure}
\begin{figure}
	\centering
	\includegraphics[width=0.44\textwidth]{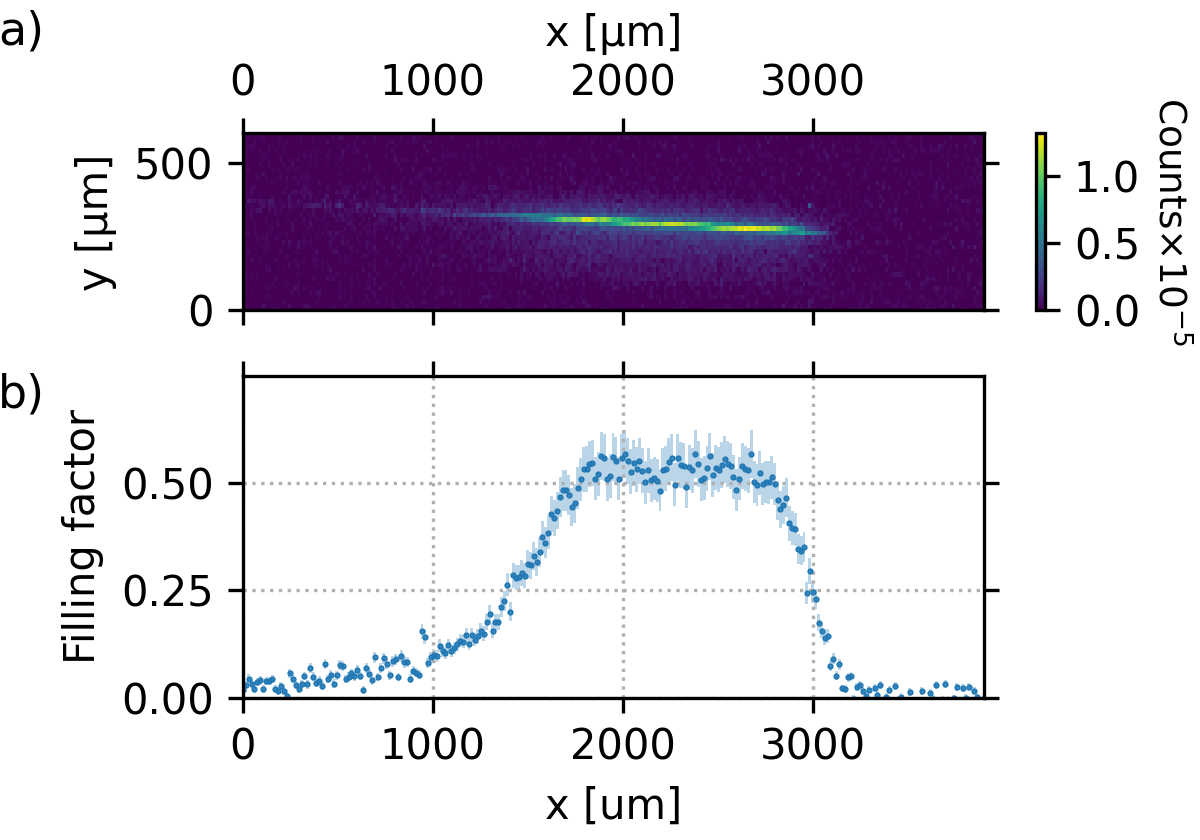}
	\caption{a) EMCCD-fluorescence image recorded while probing with fiber-guided light. b) In conjunction with an independent atom number measurement, we determine the local filling factor of the trapped ensemble. We observe a flat-topped atom distribution with a maximum filling factor of about 50\%~, which indicates trap operation in the collisional blockade regime.}
    \label{fig:FillingRatio}
\end{figure}

We start by characterizing the loading dynamics of the nanofiber trap during the RPGC and $\Lambda$GM phases. Figure~\ref{fig:Loading} shows the measured OD as a function of the duration of these two cooling stages. During the RPGC, the OD remains approximately zero for the first 15~ms, which we attribute to the atoms being still too hot for being loaded into our comparatively shallow traps when ramping down the magnetic fields. After this initial period, the measured OD increases, saturating at OD = $\SI{20.4(1.2)}{}$. The measured data are well described by a saturating exponential with a time constant of $\tau_{\mathrm{RPGC}} = \SI{9.4(1.6)}{\ms}$.  
After \SI{45.0}{\ms}, we switch off the RPGC beams and start the $\Lambda$GM phase. As shown by the blue data points in Fig.~\ref{fig:Loading}, a pronounced increase in the optical depth is observed, saturating at $\mathrm{OD} = \SI{125(4)}{}$. This corresponds to a 6-fold enhancement of the OD compared to the RPGC phase alone. The data are again well described by a saturating exponential with a time constant of $\tau_{\Lambda\mathrm{GM}} = \SI{18.0(1.1)}{\ms}$. For a fixed current in the Cesium dispensers, this sequence, combining RPGC and $\Lambda$GM, yields the highest OD. In particular, attempting to load the nanofiber trap using only $\Lambda$GM (i.e., without the initial RPGC stage) results in significantly lower OD. By increasing the Cesium dispenser current, even larger OD can be obtained without any modifications to the experimental sequence (see Supplementary Material).
These measurements were performed at the experimentally found optimal two-photon detuning, $\Delta_{2\mathrm{P}} = \SI{-5.9(0.6)}{\kHz}$, yielding the largest OD after loading by the $\Lambda$GM. This was determined by fitting a Gaussian to the data as shown in the inset of Fig.~\ref{fig:Loading}. As expected, the value of $\Delta_{2\mathrm{P}}$ is very close to the two-photon resonance, confirming that optimal loading occurs when atoms are in the velocity-selective dark state of the $\Lambda$GM. We attribute the observed offset from $\Delta_{2\mathrm{p}} = 0$ to residual differential light shifts induced by the trapping light fields and possible residual magnetic fields. \\ \indent
As depicted in Fig.~\ref{fig:Loading}, we observe a saturation of the loaded OD when applying the $\Lambda$GM for loading durations exceeding $\SI{50}{\ms}$. To understand this observation, we proceed by estimating the filling factor, i.e., the probability of finding a single atom in each trap site. This is performed by first measuring the total number of trapped atoms~\cite{Beguin2014} and then estimating the line density of the trapped atoms along the nanofiber by imaging the atomic fluorescence using a electron-multiplying charge-coupled device (EMCCD)~\cite{Pache_2025} as seen in Fig.~\ref{fig:FillingRatio} a). With these two measurements at hand, we can calculate the filling factor, considering that the distance between two trap sites is given by $\lambda_{\mathrm{red}} / (2 n_{\mathrm{eff}}) =\SI{434}{\nm}$, where $n_{\mathrm{eff}} = 1.08$ is the effective refractive index of the optical mode. Furthermore, there are two diametral one-dimensional arrays of trapping sites above and one below the nanofiber as seen in Fig.~\ref{fig:Setup}. As apparent from Fig.~\ref{fig:FillingRatio} b), we observe a roughly constant line density of the trapped atoms over a region of 1~mm. We calculate that, in this region, the filling factor for each trapping site is about 50\%, i.e., the maximum allowed in the collisional blockade regime~\cite{schlosser_collisional_2002-1}. Beyond this point, no further increase in the number of loaded single atoms occurs in the center of our ensemble. Thus, in order to reach an even larger OD, the size of the atomic cloud would have to be increased. This result is further substantiated by measuring the traps' atom loading and loss rates, as detailed in the Supplementary Material. We note that on the right-hand side in Fig.~\ref{fig:FillingRatio}, the up-tapered region of the tapered optical fiber begins. There, the intensities of the evanescent fields rapidly decreases and stable trapping is no longer possible. We note that it has been reported previously that the collisional blockade regime has been reached in a nanofiber cold atom setup solely based on the shape of the atomic fluorescence signal~\cite{Vetsch2012}, but the number of trapped atoms and consequently the filling factor was not measured.

\section{Cooling of trapped atoms}
\begin{figure}
	\centering
	\includegraphics[width=0.5\textwidth]{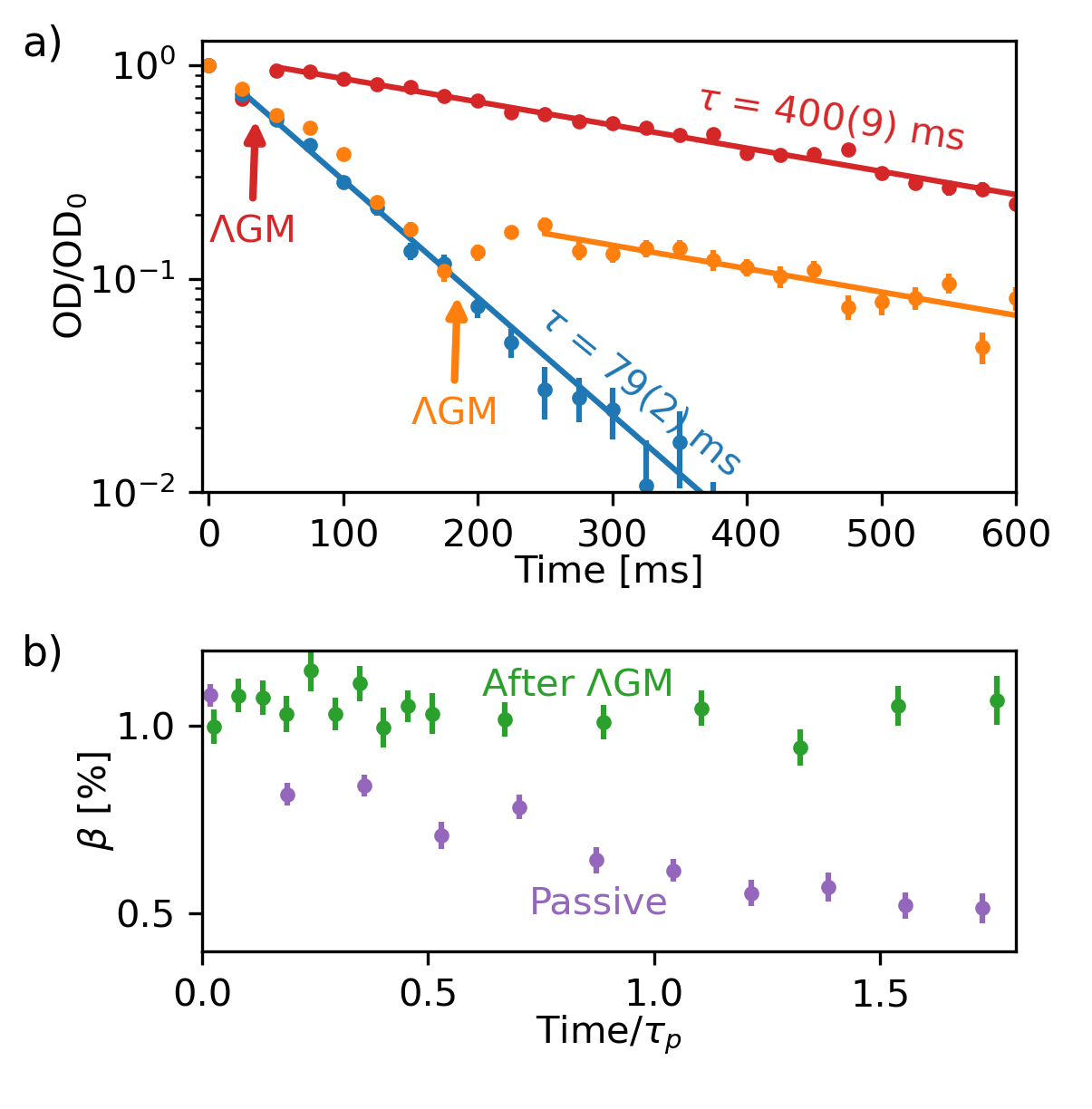}
	\caption{ a) Measurement of the storage time of the trapped atoms after loading: The blue curve shows a measurement of the passive storage time of $\tau_{p} = \SI{79(2)}{\ms}$ of the atoms in the nanofiber-guided dipole trap. The red curve shows the active storage time of $\tau_{a} = \SI{400(9)}{\ms}$ when the atoms are cooled by the $\Lambda$GM after $\SI{45}{\ms}$ of delay. For the orange curve, the $\Lambda$GM-cooling light fields are switched on after a delay of $\SI{195}{\ms}$. b) Measurement of the atom-light coupling strength $\beta$ for different holding times normalized by the passive storage time $\tau_{p}$: For a non-cooled atomic ensemble (purple) $\beta$ decreases due to heating of the atoms. When applying the $\Lambda$GM-cooling light fields (green) the atom-light coupling strength is constant with its initial value of $\beta \approx 1\% $.}
    \label{fig:Cooling}
\end{figure}

We now investigate the effect of the $\Lambda$GM light fields on the atoms that are already stored in the nanofiber trap. Because the atoms are spatially confined, the $\Lambda$GM beams are expected to induce not only polarization-gradient cooling but also EIT-cooling~\cite{Morigi2000, Phatak_2024}. This additional cooling mechanism exploits the strongly asymmetric Fano absorption profile that can be realized in a $\Lambda$-type scheme when the control beam is far-detuned with respect to the atomic resonance. In the following, we define the two-photon detuning $ \Delta_{\mathrm{2P,EIT}}$ as zero, when the the probe and coupling light fields resonantly couple the hyperfine ground states with the same motional quantum number. In this case, the ensemble becomes transparent for the probe field for transitions that do not change the motional state of the atoms, while the probe field is most absorbed for the transitions in which one quantum of motion is removed from the system. We begin our investigation of the cooling mechanism by measuring the passive storage time of the atoms in the nanofiber traps. To do so, we turn off all molasses beams after loading and then measure the OD as a function of the holding time in the trap as depicted in Fig.~\ref{fig:Cooling} a). By fitting an exponential decay to our data, we determine a storage time of $\tau_{p} = \SI{79.0(1.5)}{\ms}$. We then repeat this experiment, but after a certain delay we turn the $\Lambda$GM laser beams on again. To apply $\Lambda$GM-cooling exclusively on atoms in the nanofiber trap and avoid reloading of empty trap sites, this delay time must be long enough for the cloud of surrounding, non-trapped atoms to drop sufficiently below the nanofiber. In our setup, this takes about $\SI{20}{\ms}$. Figure~\ref{fig:Cooling} a) shows the results of these operation for delay times of 45 and $\SI{175}{\ms}$  (red and orange data points, respectively).
Interestingly, we observe that in both cases, right after turning on the cooling beams, the OD first increases on a timescale of few tens of ms and then decays with a time constant of $\tau_{a} = \SI{400(9)}{\ms}$. The latter is a five-fold increse with respect to the passive case. We note that in our comparatively shallow traps, we have not succeeded to increase the storage time of the trapped atoms using RPGC.

To interpret these results, we measure the coupling strength, $\beta$, as function of the holding time in the nanofiber trap with and without the $\Lambda$GM cooling light as seen in Fig.~\ref{fig:Cooling} b). For a non-cooled atomic ensemble (purple dots), we observe a steady decrease of $\beta$ over time. We attribute this phenomenon to heating of the atoms in the traps, which changes the spread of the atoms' distances and hence the average coupling between the atoms and the nanofiber~\cite{piotrowski2026}. Instead, when performing $\Lambda$GM-cooling during the holding time (green dots), no change in $\beta$ is observed, indicating that in this case the atoms remain at the same temperature as right after the loading. 

\section{Fiber-guided EIT-cooling }

In our nanofiber-based trap, atom loss is ultimately caused by heating in the radial direction. Interestingly, the radially decaying intensity profile of the evanescent guided light field enables the detection of the radial motional sidebands of nanofiber-trapped atoms~\cite{Ostfeldt2017}. This suggests that direct cooling of the radial motional degree of freedom using exclusively nanofiber-guided light should be possible. This is in contrast to conventional cooling schemes based on paraxial light fields, where laser cooling along directions orthogonal to the propagation direction is typically prohibitively inefficient.

\begin{figure}
	\includegraphics[width=0.5\textwidth]{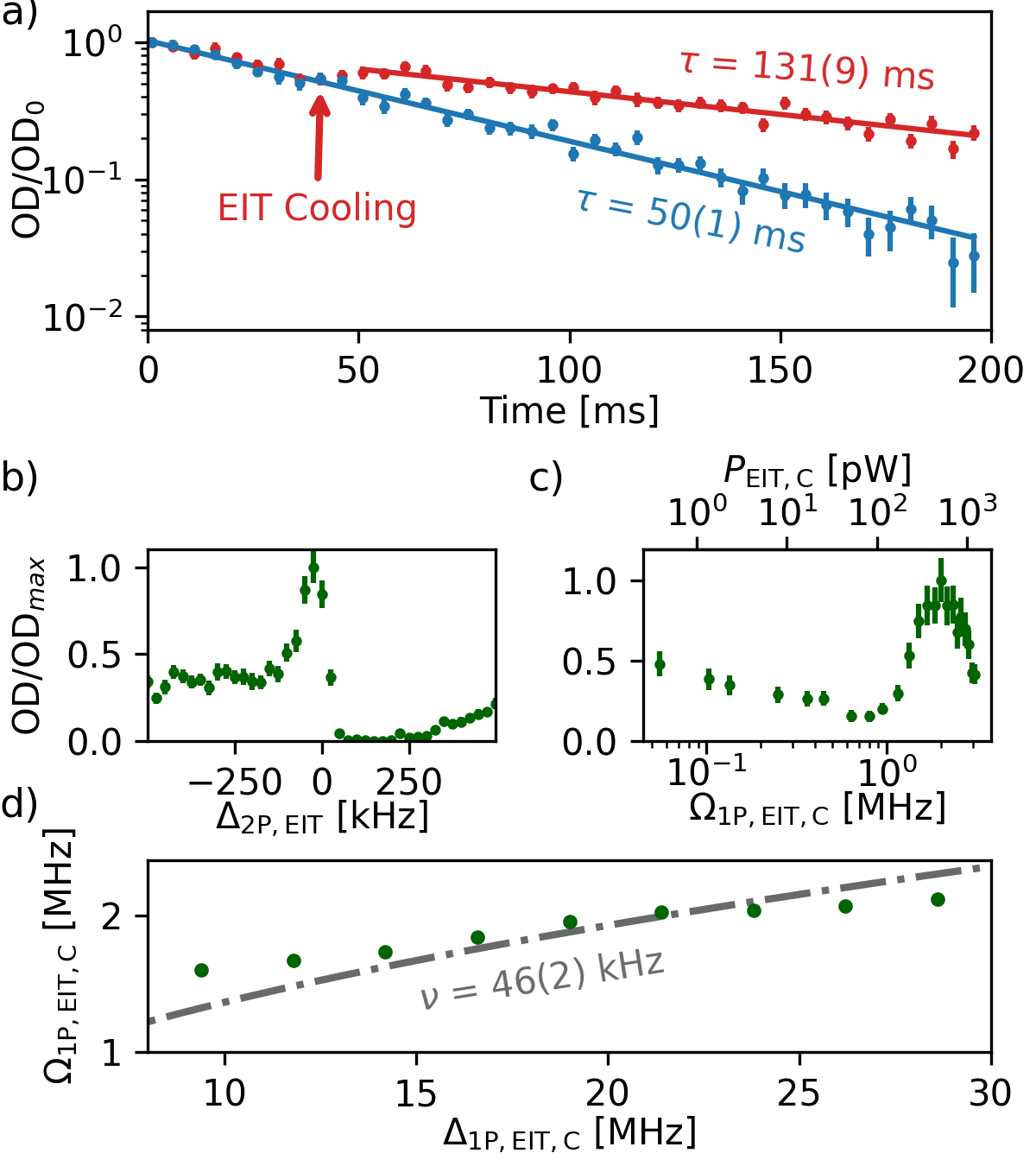}
	\caption{a) Measurement of the storage time of the trapped atoms after loading when applying no-cooling (blue) and the EIT-cooling light fields through the fiber (red). The blue curve is an exponential fit that yields a passive storage time of $\tau_{p} = \SI{50(1)}{\ms}$ of the atoms in the nanofiber-guided dipole trap. The red curve yields an active storage time of $\tau_{a} = \SI{131(9)}{\ms}$ when the atoms are cooled with the fiber-guided EIT-cooling light fields after $\SI{41}{\ms}$ of delay. b) We observe a Fano-type resonance in OD after a cooling duration of 20 ms when the EIT light fields are scanned across the two-photon resonance  $ \Delta_{\mathrm{2P,EIT}}  $ c) We find the maximum OD after applying EIT-cooling for a power of the EIT-cooler ligh field of about \SI{500}{\pico\watt}. d) By repeating this measurement for a range of one-photon detunings, $\Delta_{\mathrm{1P, EIT,C}}$, we determine $\Omega_{\mathrm{1P, EIT,C}}$ and subsequently a trap frequency of $\nu = \SI{46(2)}{\kHz}$.}
    \label{fig:EIT_Cooling}
\end{figure}

To test cooling with guided light fields, we establish a $\Lambda$-system on the D2 transition, as depicted in Fig.~\ref{fig:Setup} b), by introducing two additional phase-coherent guided light fields that co-propagate through the nanofiber. Figure~\ref{fig:EIT_Cooling} a) shows the measured storage time with and without the presence of the EIT-cooling via guided beams for a one-photon detuning of the strong coupling laser, $\Delta_{\mathrm{1P, EIT,C}}$ = $2.5\Gamma_{\mathrm{D2}}$. Again, two effects are visible: first, the storage time of the atoms in the trap is extended by a factor 2.6 and, second, a slight increase in the measured OD is observed when the cooling beam are switched on. This suggests that EIT-cooling via guided beams is also able to recover the initial value of the coupling strength, $\beta$, by cooling the atoms within the traps.

To better understand the cooling mechanism in this case, we show in Fig.~\ref{fig:EIT_Cooling} b), a measurement of the OD after a cooling time of 20 ms for different two-photon detunings $\Delta_{\mathrm{2P, EIT}}$. The result shows the Fano resonance characteristic of the EIT-cooling process, in which negative values of $\Delta_{\mathrm{2P, EIT}}$ results in cooling, while positive values result in heating. We find that the maximum OD is obtained for $\Delta_{\mathrm{2P,EIT}}$ around \SI{-25}{\kHz}, i.e., very close to two photon resonance. This offset can be attributed to residual differential light shifts induced by the trapping light fields and residual magnetic fields. The maximum cooling rate is expected when the absorption feature of the Fano resonance peaks at a two-photon detuning, $\Delta_{\mathrm{2P,EIT}}$, equal to the trap frequency, $\nu$. This occurs when the one-photon detuning, $\Delta_{\mathrm{1P,EIT,C}}$, and the Rabi frequency of the coupling beam, $\Omega_{\mathrm{1P,EIT,C}}$, satisfy the relation~\cite{Morigi2000}:
\begin{equation}
\frac{1}{2} \sqrt{ \Delta_{\mathrm{1P,EIT,C}}^2 + \Omega_{\mathrm{1P,EIT,C}}^2 } - | \Delta_{\mathrm{1P,EIT,C}} | = \nu .
\label{eq:dressing}
\end{equation}
We investigate this conditions by measuring the obtained OD after the application of EIT cooling when changing the Rabi frequency of the coupling beam, $\Omega_{\mathrm{1P, EIT, C}}$. As depicted in Fig.~\ref{fig:EIT_Cooling} c), for $\Delta_{\mathrm{1P, EIT,C}}$ = $2.5\Gamma_{\mathrm{D2}}$ we observe a maximum of the OD for a power of the EIT-cooler light field of \SI{500}{\pico\watt}, which correspond to 30\% of the saturation power. We repeat this measurement for different values of $\Delta_{\mathrm{1p, EIT,C}}$ and determine $\Omega_{\mathrm{1P, EIT,C}}$  at the maximum OD, as depicted by the green dots in  Fig.~\ref{fig:EIT_Cooling} d). We find that our data fits the prediction of Eq.~(\ref{eq:dressing}) for a trap frequency of \SI{46(2)}{\kHz}. This is in reasonable agreement with the calculated trap frequencies in our experiment of \SI{55}{\kHz} in the radial, \SI{32}{\kHz} in the azimuthal and \SI{81}{\kHz} in the axial direction, considering that Eq.~(\ref{eq:dressing}) was derived in the context of a one-dimensional model~\cite{Morigi2000}. This result confirms that EIT cooling is the leading mechanism in the case of our fiber-guided cooling technique.\\ 

\section{Conclusions}
In conclusion, our results establish $\Lambda$GM and related cooling techniques as important tools for nanophotonic interfaces with laser-cooled atoms. We demonstrate that these methods can substantially enhance the loading of nanophotonic traps, reaching filling factors that are currently limited by collisional blockade. Importantly, since $\Lambda$GM operates with blue-detuned light, it could in principle be used to achieve filling factors beyond the collisional blockade limit via tuned light-assisted collisions~\cite{Gruenzweig2010, brown_gray-molasses_2019}. For this reason, implementing $\Lambda$GM in nanophotonic cold-atom interfaces could prove crucial for investigating collective radiative phenomena that require high filling factors, such as selective radiance~\cite{AsenjoGarcia2017, Pache_2025}.
This regime, however, was not explored in the present work. In fact, this would require the trap depth $U \gtrsim \hbar\Gamma_{D1}$, which is not the case in our experiment~\cite{brown_gray-molasses_2019}. 
All of our results have been obtained in comparatively shallow trapping potentials requiring minimal optical power. This an important advantage for integrated nanophotonic platforms, where optical absorption and heat dissipation impose critical constraints.
In this context, we also demonstrate that efficient EIT-cooling using exclusively fiber-guided beams requires power as low as few hundred pW.
Looking ahead, it would be highly desirable to extend the demonstrated techniques, e.g., for directly loading atoms into nanophotonic traps, without the need for externally applied cooling laser beams.

\section*{Acknowledgment}
We thank Hector Letellier for helpful discussions and Johannes Piotrowski for assistance with the EMCCD setup. We acknowledge funding by the Alexander von Humboldt Foundation in the framework of the Alexander von Humboldt Professorship endowed by the Federal Ministry of Education and Research, and SuperWave (ERC Grant No. 101071882).
\bibliography{gray.bib}

\clearpage

\section*{Supplementary Material}

\subsection{Experimental sequence}

\begin{figure}
	\centering
	\includegraphics[width=0.47\textwidth]{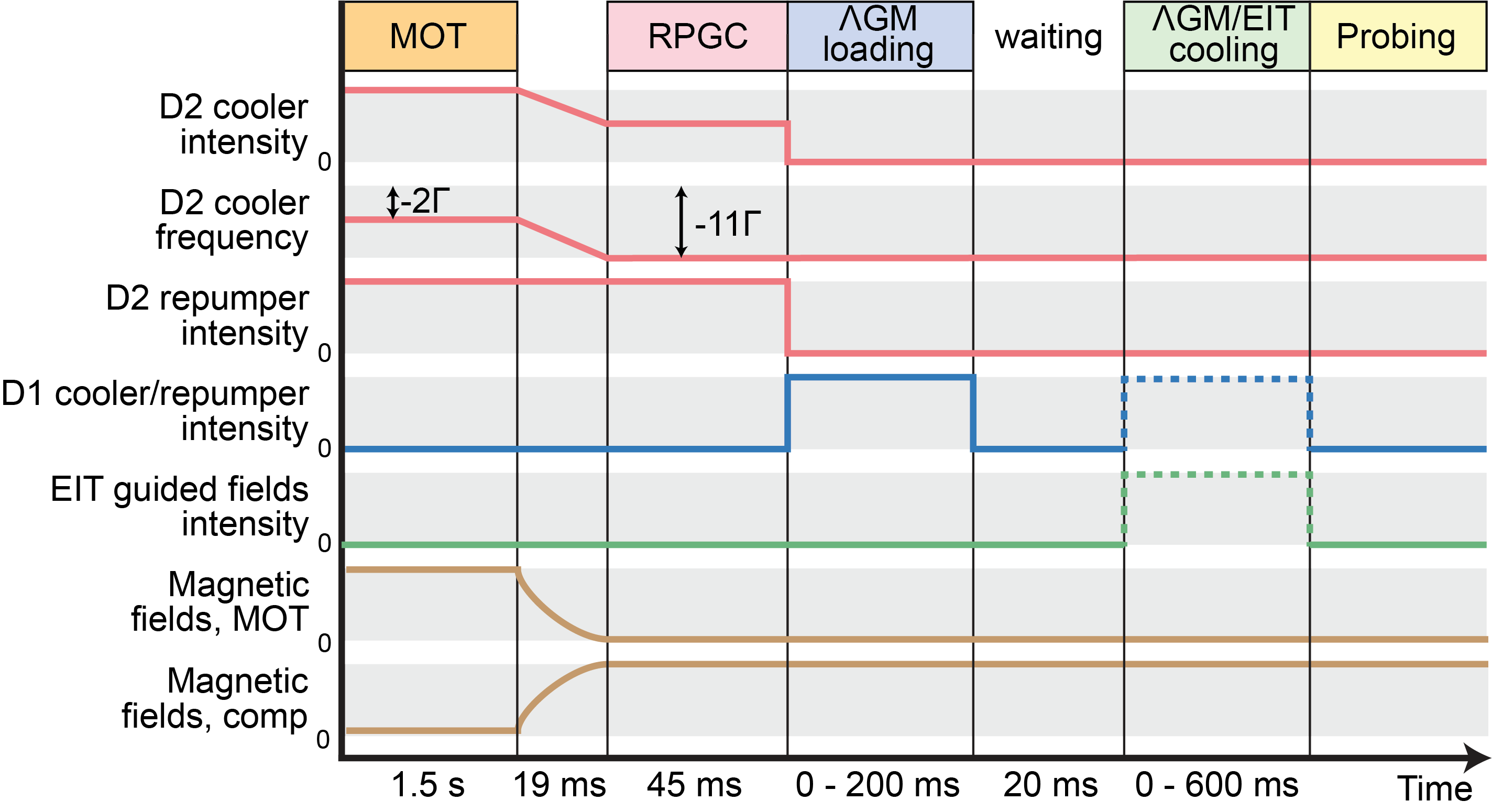}
	\caption{Diagram of our experimental sequence. The dashed lines indicate that we perform either cooling with the D1 cooling/repumper beam or via the EIT guided fields.}
    \label{fig:Sequence}
\end{figure}

The experimental sequence used to performed the measurements is shown in Fig.~\ref{fig:Sequence}.
We begin by preparing a cloud of cold Cesium atoms in a magneto-optical trap (MOT) for a duration of 1.5~s. During this stage, the atoms are cooled using a laser, red-detuned by $2\Gamma_{D2}$ from the $F=4 \rightarrow F^{\prime}=5$ D$_2$ cycling transition, hereafter referred to as the D2-cooler. Here, $\Gamma_{D2}$ denotes the natural linewidth of this transition. An additional D2-repump laser, resonant with the $F=3 \rightarrow F^{\prime}=4$ transition, is employed to prevent atoms from accumulating in the $F=3$ hyperfine ground state. The pairs of counter-propagating light fields are $\sigma^{+}$ and $\sigma^{-}$ polarized.

After loading the MOT, in a 19 ms long transition phase, we ramp the magnetic field gradient down to zero, gradually reduce the intensity of the D2 cooling beam and increase its detuning to $11\Gamma_{D2}$.
We then perform a first stage of red-detuned polarization-gradient cooling (RPGC) lasting 45~ms. At the end of this phase, the atomic molasses reaches a temperature of approximately 5.5~$\mu$K, as measured via time-of-flight. 

Next, we implement $\Lambda$-enhanced gray molasses ($\Lambda$GM) cooling. For this step, the D2 cooling and repumping lasers are switched off and two light fields driving a $\Lambda$ system are applied on the D1 transition between the two hyperfine ground states and the $6\text{P}_{1/2},\,F^{\prime}=4$ excited state as depicted in Fig.~\ref{fig:Setup}b). Again, the counter-propagating light fields of $\Lambda$GM are $\sigma^{+}$ and $\sigma^{-}$ polarized. The first field, referred to as the D1-cooler, is blue-detuned by $\Delta_\mathrm{1P} = 4.5\Gamma_\mathrm{D1}$ from the $F=4 \rightarrow F^{\prime}=4$ transition. We produce a phase-coherent second light field by using an electro-optic modulator (EOM QUBIG PM-Cs\_9.2) which creates sidebands around the cooling laser at a frequency corresponding to the the hyperfine ground state splitting of Cesium, with a sideband to carrier ratio of 1 to 4. The parameters for loading the atoms using $\Lambda$GM (one-photon detuning $\Delta_\mathrm{1P}$, two-photon detuning $\Delta_\mathrm{2P}$ and sideband-to-carrier ratio) were optimized to yield the highest optical depth (OD). We use an atomic clock (Safran mRO-50) to precisely define the modulation frequency sent to the EOM. The $\Lambda$GM stage lasts up to 200~ms.

For measurements involving cooling of the already trapped atoms, all molasses beams are switched off for 20~ms after loading, allowing the cold atomic cloud to fall below the nanofiber. Afterwards, we apply the $\Lambda$GM as described above. 

For the measurements that demonstrate EIT cooling using exclusively guided field, we launch in the nanofiber a laser beam that is $2.5\Gamma_\mathrm{D2}$ blue-detuned with respect to the $F=4 \rightarrow F^{\prime}=4$ hyperfine transition of the Cesium D2-line with a power of about \SI{500}{\pico\watt}. In this case, the sidebands are generated using a fiber-based phase modulator (Exail NIR-MPX800-LN-10) and the carrier to sideband ratio is 13 to 1.
The transverse polarization of the EIT-cooling light fields are aligned to be orthogonal to the plain containing the trapped atoms.

Finally, during the probing phase, we characterize the relevant properties of the atomic ensemble, such as the optical depth and the number of trapped atoms, as described in Section C.

\subsection{Temperature of the untrapped atom cloud after $\Lambda$GM}

\begin{figure}
	\centering
	\includegraphics[width=0.44\textwidth]{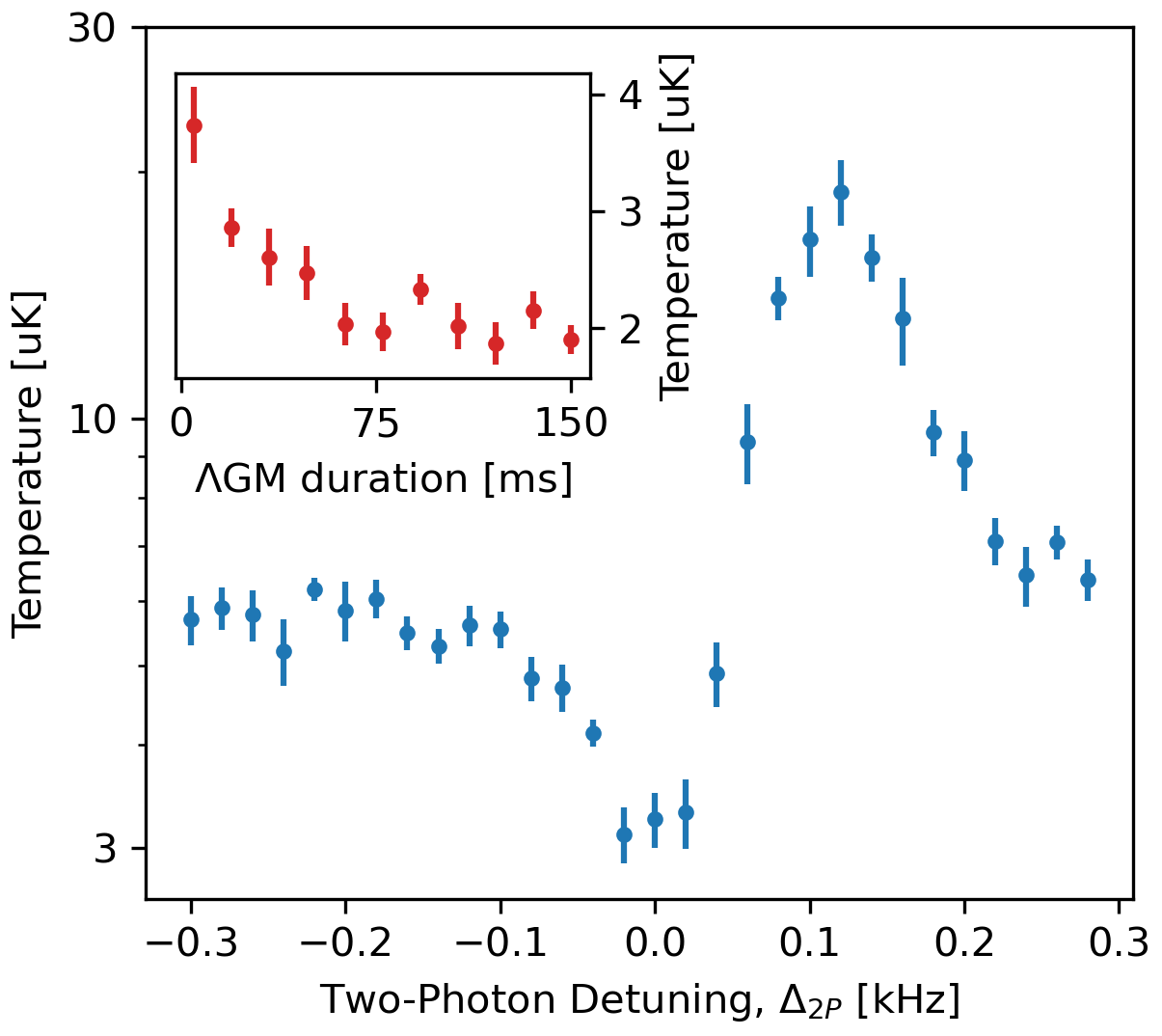}
	\caption{Measurement of the temperature of the cloud of Cesium atoms after \SI{3}{\ms} of $\Lambda$GM cooling as a function of the two-photon detuning, $\Delta_{\mathrm{2P}}$. Inset: measurement of the temperature of the atom cloud as a function of the duration of the $\Lambda$GM cooling stage for $Delta_{\mathrm{2P}} = 0$.}
    \label{fig:Fano}
\end{figure}

Figure~\ref{fig:Fano} shows the temperature of the untrapped Cesium atom cloud after $\Lambda$GM cooling, measured via time-of-flight as a function of the two-photon detuning, $\Delta_{2P}$. For these measurements, the cooling duration was fixed at \SI{5}{\ms}. We observe the characteristic Fano-like dependence of the temperature expected for this cooling scheme, with a minimum at zero two-photon detuning, where the atoms predominantly occupy the velocity-dependent dark state. The inset of Fig.~\ref{fig:Fano} presents the molasses temperature as a function of the $\Lambda$GM duration. For a cooling time of \SI{150}{\ms}, we measure a final temperature of \SI{1.9(1)}{\micro\K}.

\subsection{Measurements of optical depth, OD, and atom number $N$}

\begin{figure}
	\centering
	\includegraphics[width=0.44\textwidth]{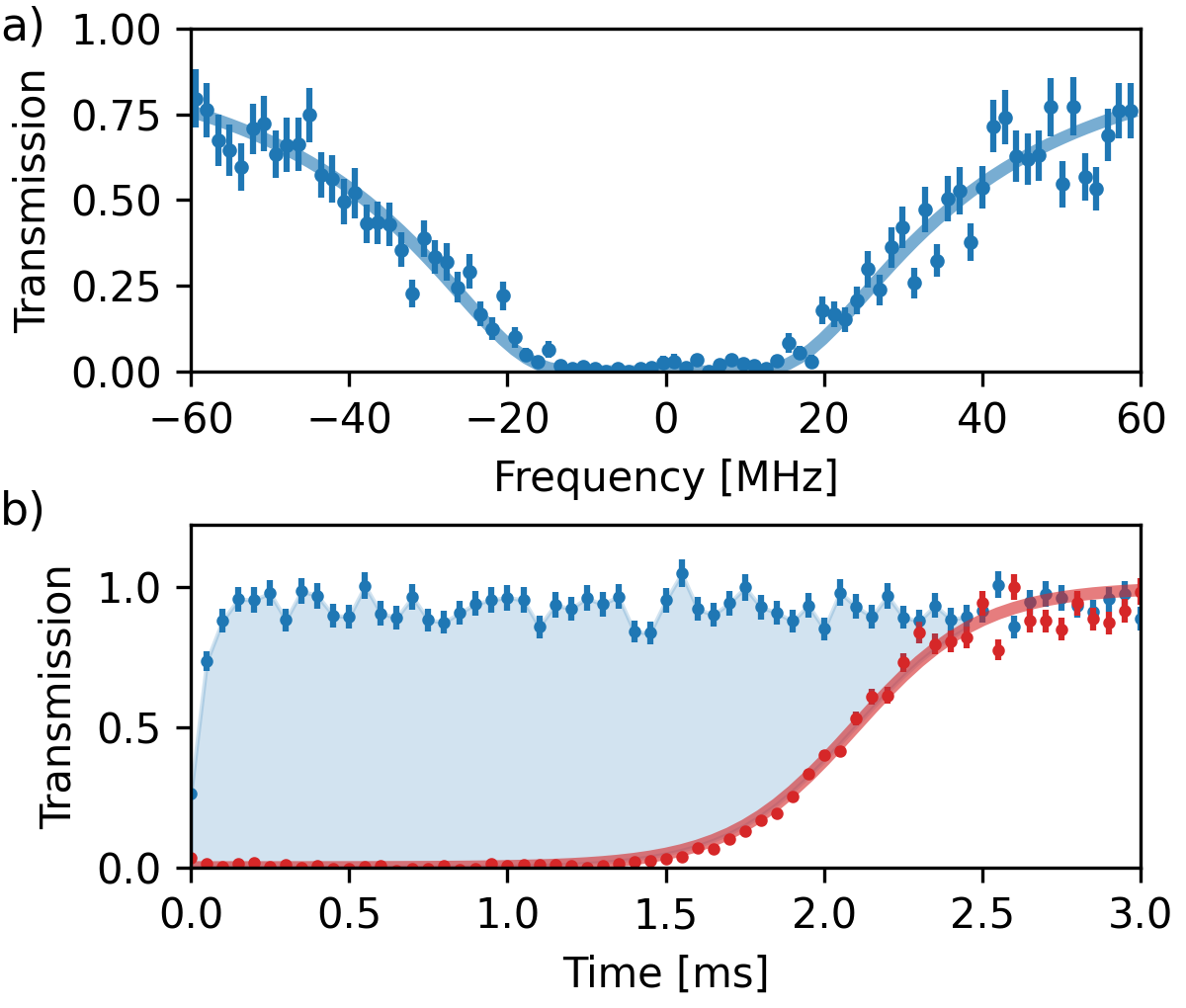}
	\caption{a) Typical transmission spectroscopy measurement used to determine the optical depth (OD). A fit using a saturated Lorentzian model yields an OD of 144(4). b) Time-dependent transmission measurement obtained using the depumping technique (red points). By fitting the transmission dynamics with the model developed in Ref.~\cite{Beguin2014}, we extract an atom number of $N = 4360 \pm 230$. The blue points show a reference measurement performed in the absence of trapped atoms.}
    \label{fig:SM_ODandN}
\end{figure}

We estimate the OD of the nanofiber-trapped atoms by measuring the absorption of the atomic ensemble while scanning a weak probe laser across the $F=4 \rightarrow F^{\prime}=5$ hyperfine transition of the D2 line. To obtain the OD, we fit the measured transmission spectrum with a saturated Lorentzian. The probe power is set to approximately $10^{-2} P_{\mathrm{sat}}$, where $P_{\mathrm{sat}}$ is the saturation power for this transition. Here, the transverse polarization of the probe beam is parallel to the plain containing the trapped atoms. The total duration of the probe scan is 500~$\mu$s. A typical transmission spectroscopy is shown in Fig.~\ref{fig:SM_ODandN}~a) with an OD of 144(4). 

The atom number, $N$, is determined using the method developed in Ref.~\cite{Beguin2014}. In our implementation, a weak probe beam resonant with the $F=4 \rightarrow F^{\prime}=4$ hyperfine transition of the Cesium D2 line is sent through the nanofiber with an optical power of \SI{1.3}{pW}. The probe polarization is chosen to be transverse and orthogonal to the plane containing the trapped atoms. To determine the atom number, we record the transmission of a boxcar shaped probe pulse as a function of time. A representative transmission trace is shown in Fig.~\ref{fig:SM_ODandN}b) (red points), together with a reference measurement taken without trapped atoms (blue points). We observe that the initial transmission, close to zero due to the large optical depth, gradually increases toward unity as the atoms are optically pumped into the other $F=3$ hyperfine ground state.
Specifically, the atom number is extracted by fitting our transmission measurement to the model developed in Ref.~\cite{Beguin2014}. For the dataset in Fig.~\ref{fig:SM_ODandN}b) we obtain $N = 4360 \pm 230$ atoms. Alternatively, $N$ can be also determined by directly counting the missing photons in the transmission signal, i.e., by evaluating the area of the blue shaded region in Fig.~\ref{fig:SM_ODandN}b), taking into account the branching ratios to the two hyperfine ground states. Both methods agree within our experimental uncertainty.

\subsection{Characterization of atom loading and loss rates}

In order to further compare the $\Lambda$GM and the RPGC loading techniques, we estimate for these two cooling schemes the one-atom loss rate, $\gamma$, the one-atom loading rate, $R$, and the normalized two-body loss rate, $\kappa / 4V$. Here, $V$ is the trap volume and $\kappa = \SI{10e-10}{\centi\meter\cubed  \per\second}$ is the two-body loss rate of Cesium~\cite{Ueberholz_2000}.

We determine the one-atom loss rate as $\gamma = 1/\tau$, where $\tau$ is the storage time of the trapped atoms while cooling them using $\Lambda$GM or RPGC. We obtain $\gamma_{\mathrm{\Lambda GM}} = \SI{2.50(1)}{\per\second}$ and $\gamma_{\mathrm{RPGC}} = \SI{50(1)}{\per\second}$.

We estimate the loading rate, $R$, from the initial rate at which the number of trapped atoms increases, $\frac{dN}{dt}|_{t=0}$, right after switching on the RPGC or $\Lambda$GM beams (see data in Fig.~\ref{fig:Loading}, in which we assumed that $N = \mathrm{OD}/(4 \beta)$, with $\beta =1 \%$).  The loading rate, $R$, is then given by $\frac{dN}{dt}|_{t=0}/N_{traps}$, where $N_{traps}$ denotes the number of trapping sites available along the nanofiber given the size of our atom cloud, which we estimate to be \SI{1.5}{\mm} (see Fig.~\ref{fig:FillingRatio}). With that, we can determine $R_{\mathrm{RPGC}} = \SI{7(1)}{\per\second}$ and $R_{\mathrm{\Lambda GM}} = \SI{25(2)}{\per\second}$.

We calculate, $V$, by computing the full width at half maximum, $d$, of the trap potential in each direction in cylindrical coordinates. We get $d_r = \SI{270}{\nano\meter}$, $d_{\theta} = \SI{0.87}{\radian}$ and $d_z = \SI{150}{\nano\meter}$. We can than calculate the trapping volume based on Ref.~\cite{Kuppens2000} as
\begin{equation}
V = d_r d_z d_{\theta}r \cdot \ln\left( \frac{1}{1-\eta}\right) \sqrt{\frac{\eta}{1-\eta}}.
\label{Eq:volume}
\end{equation}
Here $\eta = k_B T/|U_0|$ is the ratio between the trap depth $U_0$ and the temperature of the atoms multiplied by Boltzmann's constant $k_B$. Inserting our trap parameters into Eq.~\ref{Eq:volume}, we obtain $V = \SI{1.09e-16}{\centi\meter\cubed}$.

These results are summarized in Table~\ref{tab:loading} and show that for $\Lambda$GM the atom loading is indeed limited by collisional blockade as the following condition is satisfied~\cite{schlosser_collisional_2002-1}: $ \gamma/2 < R < \kappa / 4V$.
Instead, in our experiment, RPGC molasses operates in the regime of weak loading in which the mean atom number per trap site is given by $\langle N \rangle = R_{\mathrm{RPGC}} / \gamma_{\mathrm{RPGC}}$~\cite{schlosser_collisional_2002-1}. 

\begin{table}[]
\centering
\setlength{\tabcolsep}{20pt}
\renewcommand{\arraystretch}{1.2}
\begin{tabular}{lcc}
\hline\hline
 & $\Lambda$GM & RPGC \\
\hline
$R$ (s$^{-1}$)           & $25 \pm 2$              & $7 \pm 1$              \\
$\gamma$ (s$^{-1}$)      & $2.50 \pm 1$            & $50 \pm 1$              \\
\hline\hline

\end{tabular}
\caption{Atom loading $R$ and loss rates $\gamma$ for $\Lambda$GM and RPGC cooling schemes. In our nanofiber-based dipole trap the normalized two-body loss rate $\kappa / 4V$ is $\SI{2.2e6}{\per\second}$.}

\label{tab:loading}
\end{table}

\subsection{Fluorescence imaging of the nanofiber trapped atoms}
We image the trapped atomic ensemble by collecting their fluorescence when illuminated through the nanofiber. In order to obtain ensemble with an high atom number, we prepare a cloud of cold atoms by loading the MOT for \SI{1.2}{\second}. Afterwards, we are using the RPGC molasses for \SI{60}{\milli \second} an, subsequently, the $\Lambda$GM for \SI{190}{\milli \second} to load the atoms into the nanofiber-guided dipole traps. For probing the atoms, we use the EIT-cooling light fields on the D2 transition for which we choose a detuning of about $7.3 \Gamma_{D2}$. We image the atom trapped on the nanofiber perpendicular to the propagation direction of the light onto an EMCCD (Andor iXon Ultra). For the image seen in Fig.~\ref{fig:FillingRatio}, we repeat the imaging sequence 3000 times while using an integration and probing time of \SI{23}{\milli \second}.

\end{document}